\documentstyle[preprint,pra,aps]{revtex}
\tightenlines
\begin{document}
\draft
\title{\bf Atomic  optical clocks and
search for variation of the fine structure constant.}
\author{V.A.Dzuba$^*$ and V.V.Flambaum}
\address{School of Physics, University of New South Wales, 
Sydney 2052,Australia}
\date{\today}
\maketitle

\begin{abstract}
Theories unifying gravity and other interactions suggest the
possibility of spatial and temporal variation of physical
``constants''. Accuracy achieved for the atomic optical frequency
standards (optical clocks) approaches the level when possible
time evolution of the fine structure constant $\alpha$ can be
studied by comparisons of rates between clocks based on
different atomic transitions in different atoms. The sensitivity
to variation of  $\alpha$ is due to relativistic corrections
which are different in different atoms ($\sim Z^2\alpha^2$).
We have calculated the values of the relativistic energy shifts in In~II,
Tl~II, Ba~II and Ra~II which all can be used as atomic optical clocks.
The results are to be used to translate any
change in the clock's rate into variation of $\alpha$.
\end{abstract} 
\vspace{1cm}

Possible variations of the fundamental physical constants are suggested
by unified theories, such as string theory and $M$ theory (see, e.g.
\cite{Marciano,Barrow,Damour94}).
A number of works have been done in last few years in an attempt to
find an experimental evidence of any space-time variation of the
fine structure constant $\alpha$.
The search goes mostly in two ways. 
One is based on the analysis of the absorption spectra of
distant quasars. Comparing the spectra of atoms or ions in distant gas
clouds which intersect the sight lines towards the quasars with the
laboratory spectra allows to put bounds on the space-time variation of
$\alpha$.
Another way uses precise atomic clocks in laboratory measurements.
Different atomic transitions depend differently on the fine structure
constant.
 Comparing the rates of different atomic clocks over long
period of time allows to put bounds on the local change of $\alpha$
with time.
Astrophysical measurements have a big advantage of having many-orders
of magnitude enhancement factor gained by looking into distant past. 
At present time the strongest bound on the possible space-time variation
of $\alpha$ has been obtained from the analysis of the quasar absorption
spectra. There is even an evidence that the value of $\alpha$ might be
smaller in early epochs \cite{Webb}.
However, the accuracy achieved for atomic clocks now approaches the level
where the measurements with a similar accuracy become possible.
These measurements are also important because they produce results 
which are independent of the cosmological model and any possible space 
variation of the fundamental constants.

The strongest laboratory limit on the time variation of $\alpha$ was
obtained by comparing H-maser vs Hg~II microwave atomic clocks over
140 days \cite{Prestage}. Fermi-Segre formula for the hyperfine splitting
was used to translate frequency drift into variation of $\alpha$.
This yielded an upper limit 
$\dot \alpha/\alpha \le 3.7 \times 10^{-14}/\mbox{yr}$.

Another possibility is to use optical atomic frequency standards.
These standards are based on strongly forbidden $E1$-transitions or 
$E2$-transitions
between the ground state of an atom (ion) and its close metastable
excited state. Proposed optical frequency standards include
Ca~I \cite{CaI}, Sr~II \cite{SrII}, Yb~II \cite{YbII}, Hg~II \cite{HgII},
Mg~I \cite{MgI}, In~II \cite{InII}, Xe~I \cite{XeI}, Ar~I \cite{ArI}, etc.
In contrast with the microwave frequency standards, there is no simple
analytical formula for the dependence of optical atomic frequencies on
$\alpha$. This dependence can be revealed via accurate relativistic
calculations only. In our earlier paper \cite{Dzuba99} we presented such 
calculations for Ca~I, Sr~II, Yb~II and Hg~II. We stress that relativistic
corrections can not be reduced to spin-orbit interaction. For example,
the $s$-electron level has the largest relativistic correction and
no spin-orbit interaction \cite{Dzuba99}.

Recently an experiment to measure possible time variation of $\alpha$
was proposed in Ref. \cite{Peik} by linking H and In~II optical
frequency standards.
In present work we  calculate relativistic energy shift of the clock
transition of In~II and its heavier analogue Tl~II. For the search of
variation of $\alpha$ Tl~II may be preferable since it has bigger relativistic
effects (the relative magnitude  of the relativistic corrections
 increases as $ Z^2\alpha^2$ with the nuclear charge Z).
 We also include in the calculations some other metastable states 
of both ions.
Ba~II ion had been considered as a candidate for the optical frequency
standard in Ref. \cite{BaI}. Therefore, we perform the calculations for Ba~II
and its heavier analogue Ra~II as well.

It is convenient to represent the results in the form
\begin{eqnarray}
	\omega = \omega_0 + q_1 x + q_2 y,
\label{omegaxy}
\end{eqnarray}
where 
$x = (\frac{\alpha}{\alpha_l})^2 - 1, y = (\frac{\alpha}{\alpha_l})^4 - 1$
and $\omega_0 $ is an experimental frequency of a particular transition.
To find the value of the coefficients $q_1$ and $q_2$ we have repeated the
calculations for $\alpha = \alpha_0, \alpha =
\sqrt{7/8}\alpha_0$ and $\alpha = \sqrt{3/4}\alpha_0$ and fit the results
by the formula (\ref{omegaxy}).
We started the calculations from the relativistic Hartree-Fock method.
$V^{N-1}$ approximation was used to generate a complete  set of the core and
valence basis states. Correlations between the core and valence electrons 
have been included by means of the many-body perturbation theory. 
In the case of In~II and Tl~II which both have two valence electrons 
above the closed-shell core the correlations between the valence electrons 
have been included by means of the  configuration interaction method. 
More detailed discussion of the method of calculations can be found in 
our earlier work \cite{Dzuba99}.

The obtained  values of the coefficients $q_1$ and $q_2$ as well as
 experimental
frequencies of some ``clock'' transitions in In~II, Tl~II, Ba~II and Ra~II
are presented in Table \ref{results}.
Note that
\begin{equation}
	\dot \omega \mid_{\alpha = \alpha_0} = (2q_1 + 4q_2)
	\frac{\dot \alpha}{\alpha}.  
\label{domega}
\end{equation}
The most recent and strongest limits on the time variation of $\alpha$ are
\begin{eqnarray*}
	\dot \alpha/\alpha & < & 3.7 \times 10^{-14}/\mbox{yr} \ \ \
	\mbox{Prestage {\it et al}} \ \cite{Prestage},  \\
	\dot \alpha/\alpha & < & 10^{-15}/\mbox{yr} \ \ \
	\mbox{Damour and Dyson} \ \cite{Damour},  \\
	\dot \alpha/\alpha & < & 10^{-15}/\mbox{yr} \ \ \
	\mbox{Webb {\it at al}} \ \cite{Webb},  \\
	\dot \alpha/\alpha & < & 1.9 \times 10^{-14}/\mbox{yr} \ \ \
	\mbox{Ivanchik {\it et al}} \ \cite{Ivanchik}.  \\
\end{eqnarray*}
Only first of these results is a local present-day limit on the time
variation of $\alpha$. It was obtained by comparing Hg~II and H
microwave atomic clocks as was mentioned before. 
Second result was obtained from the analysis of 
the Oklo natural nuclear reactor. The Oklo event took place in
Gabon (Africa) around $1.8 \times 10^9$ years ago. Two other results
came from the analysis of the astrophysical data and correspond to even
bigger time intervals. It is interesting to see what accuracy is needed
to improve the present-day limit on variation of $\alpha$. Substituting
$\dot \alpha/\alpha = 10^{-14}$ and $q_1$ and $q_2$ from Table \ref{results}
into formula (\ref{domega}) we can get for the In~II clock transition
\begin{equation}
	\mbox{In~II:} \ \ \dot \omega = 2.6~ \mbox{Hz/yr}\ \ \ 
	(\dot \alpha/\alpha = 10^{-14}\mbox{yr}^{-1}).
\end{equation}
Note, that the natural linewidth of the clock line of In~II is 1.1 Hz
\cite{Peik}. The frequency of the $^{115}$In~II clock transition is 
currently known to the accuracy $\sim 10^{-13}$:
\[
	\omega_0 = 1~267~402~452~914(42)~ \mbox{kHz} \cite{Peik}.
\]
However, further two orders of magnitude improvement in accuracy is
probably possible \cite{Peik1}. In fact , it is enough to measure
variation of the ratio or difference between two frequencies. 
Note, that relativistic energy shift of the $^1$S$_0 - ^3$P$_0$
transition in Tl~II is about 5 times bigger:
\begin{equation}
	\mbox{Tl~II:} \ \ \dot \omega = 12~ \mbox{Hz/yr}\ \ \ 
	(\dot \alpha/\alpha = 10^{-14}\mbox{yr}^{-1}).
\end{equation}
Relativistic effects are also big for upper metastable states of In~II
and Tl~II and for the $s -d$ transitions in Ba~II and Ra~II (see Table
\ref{results}). 
In principle, all these states can be used for atomic clocks.

We are grateful to W. Nagourney and E. Peik  for useful 
discussions.


\begin{table}
\caption{Relativistic energy shift of the $^1$S$_0 - ^3$P$_0$ clock
transition of In~II and some ground to metastable states transitions
of In~II, Tl~II, Ba~II and Ra~II (cm$^{-1}$) (see formula (\ref{omegaxy}) 
for the definition of $q_1$ and $q_2$).}
\label{results}
\begin{tabular}{cllllllrr}
 ~Z & Ion & \multicolumn{2}{c}{Ground state} & 
\multicolumn{2}{c}{Upper states} & $\omega_0$\tablenotemark[1]
 & $q_1$ & $q_2$~~ \\
\hline
 ~49 & In~II & $5s^2 $ & $ ^1$S$_0$ & $5s5p $ & $ ^3$P$_0$ & 
\dec 42275. & 2502 & 956~~ \\
     &        &         &            & $5s5p $ & $ ^3$P$_1$ & 
\dec 43349. & 3741 & 791~~ \\
     &        &         &            & $5s5p $ & $ ^3$P$_2$ & 
\dec 45827. & 6219 & 791~~ \\

 ~81 & Tl~II & $6s^2 $ & $ ^1$S$_0$ & $6s6p $ & $ ^3$P$_0$ & 
\dec 49451. & 1661 & 9042~~ \\
     &        &         &            & $6s6p $ & $ ^3$P$_1$ & 
\dec 53393. & 5877 & 8668~~ \\
     &        &         &            & $6s6p $ & $ ^3$P$_2$ & 
\dec 61725. & 14309 & 8668~~ \\
 ~56 & Ba~II & $6s $ & $ ^2$S$_{1/2}$ & $5d $ & $ ^2$D$_{3/2}$ & 
\dec 4843.850 & 5402 & 221~~ \\
     &        &         &            & $5d $ & $ ^2$D$_{5/2}$ & 
\dec 5674.824 & 6872 & -448~~ \\

 ~88 & Ra~II & $7s $ & $ ^2$S$_{1/2}$ & $6d $ & $ ^2$D$_{3/2}$ & 
\dec 12084.38 & 15507 & 1639~~ \\
     &        &         &            & $6d $ & $ ^2$D$_{5/2}$ & 
\dec 13743.11 & 19669 & -864~~ \\
\end{tabular}
\tablenotetext[1]{Moore, Ref. \cite{Moore}}
\end{table}

\end{document}